# Asymmetric Surface Plasmon Polariton Emission by a Dipole Emitter near a Metal Surface

J. P. Balthasar Mueller and Federico Capasso[*]

*School of Engineering and Applied Sciences, Harvard University, Cambridge, Massachusetts 02138, USA.*

*correspondence to capasso@seas.harvard.edu

**Abstract:** We show that the surface plasmon polariton (SPP) radiation patterns of point-dipole emitters in the vicinity of a metal-dielectric interface are generally asymmetric with respect to the location of the emitter. In particular rotating dipoles, which emit elliptically polarized light, produce highly asymmetric SPP radiation fields that include unidirectional emission. Asymmetric SPP radiation patterns also result when a dipole oscillates tilted with respect to the plane of the interface and optical losses or gains are present in the materials. These effects can be used to directionally control SPP emission and absorption, as well as to study emission and scattering processes close to metal-dielectric interfaces. Possible implementations of asymmetrically emitting SPP sources are discussed.

**Main Text:** Surface plasmon polaritons (SPPs) are tightly confined electromagnetic waves that propagate across metal surfaces at the frequencies of infrared or visible light[1]. Their strong interaction with matter close to metal interfaces has led to a variety of applications in modern biological and chemical sensing[2-4], and to new developments in photovoltaics[5] and quantum optics[6]. Because of their small mode volume and widely controllable properties, SPPs are considered promising candidates to enable on-chip information technology at optical bandwidths[7,8].

The problem of SPP radiation (or, by virtue of time-reversal symmetry, absorption[9]) by classical point-dipoles is ubiquitous in theoretical plasmonics, as it provides an accurate and efficient model for understanding a broad spectrum of phenomena in light-matter interaction. This includes the interaction with the analyte in plasmon-enhanced sensing applications[10,11], light-SPP coupling[1,12-14], SPP generation and amplification by gain media[15-17] and SPP scattering[18-20]. The analysis often distinguishes between a dipole that is oscillating in-plane and a dipole that is oscillating out-of-plane with respect to the metal surface[10,16,21,22]. In these cases the emitted fields have point-symmetry about the location of the dipole in the plane of the interface, which means that the radiated

SPP fields are invariant up to a sign change under a rotation by 180° in the plane of the metal surface. However, a radiating dipole can generally consist of components that oscillate in all three dimensions with relative phases[23,24]. Even though the emission of a dipole into free space is always symmetric[23], the emission of such dipoles into SPP modes exhibits a rich variety of radiation patterns without angular symmetry. In this letter, we investigate the SPP radiation fields of a general dipole in the vicinity of a metal surface with particular emphasis on the radiation pattern and its asymmetry. In particular, we show how rotating dipoles can give rise to highly asymmetric in-plane SPP emissions. We also show that asymmetric emission arises in the presence of optical losses when the dipole oscillates tilted with respect to the interface.

We consider an oscillating electric point-dipole that is situated on the $z$-axis at a small distance $d$ away from a metal-dielectric interface in the $x,y$ plane (Fig. 1A). The SPP fields emitted by an out-of-plane and an in-plane dipole[22] are plotted in Figs. 1B and C, respectively. The plots of the intensity are plotted on a logarithmic scale in powers of 10 to enhance visibility. The SPP fields radiated by an out-of-plane dipole have no angular dependence and are hence symmetric about the origin in both intensity and phase (Fig. 1B). An in-plane dipole gives rise to the well-known bimodal radiation field with a $\cos^2\theta$ dependence of the intensity (Fig. 1C, right), where $\theta$ is the angle with respect to the axis of the dipole. The SPP fields launched by in-plane dipoles have been directly observed experimentally with near-field probes[1], quantum emitters[25] and scattering from nanostructures[13]. In contrast to the field of the out-of-plane dipole, the phase of the field radiated by an in-plane dipole is anti-symmetric about the origin, meaning that the sign of the field-vectors changes under the transformation $\phi \to \phi + \pi$. Simultaneous oscillation both in- and out-of-plane with the metal surface then results in a superposition of symmetric and anti-symmetric fields, which generally results in non-symmetric field distributions.

We parameterize a general oscillating dipole with strength $p_0$ and real, constant frequency $\omega$ as:

$$\vec{p} = \text{Re}\left\{ \begin{bmatrix} p_x \\ p_y \\ p_z \end{bmatrix} \exp(-i\omega t) \right\} = \text{Re}\left\{ p_0 \begin{bmatrix} \sin\alpha\cos\beta\exp(i\delta) \\ \sin\alpha\sin\beta\exp(i\gamma) \\ \cos\alpha \end{bmatrix} \exp(-i\omega t) \right\} \qquad (0)$$

Where $p_x$, $p_y$ and $p_z$ are the projections of the dipole along the coordinate axes and are described by the angles $\alpha \in [0, \pi/2]$ and $\beta \in [0, \pi/2]$. The respective components oscillate with phase shifts, which are defined relative to $p_z$ by $\delta \in [-\pi, \pi]$ and $\gamma \in [-\pi, \pi]$. Far away from the emitter, the SPP electric field radiated by such a general dipole is given by[22,26]:

$$\vec{E}_n = 2\text{Re}\left\{\left[\frac{\exp(ik_{SPP}\rho)}{\sqrt{k_{SPP}\rho}}\exp(ik_{z,n}|z|)\exp(ik_{z,1}|d|)\exp(-i\omega t)\right]M\vec{v}\right.$$
$$\left. \times \left[p_x \cos\phi + p_y \sin\phi + p_z\left(\text{sgn}(d)\frac{k_{SPP}}{k_{z,1}}\right)\right]\right\} \tag{0}$$

where we used cylindrical coordinates $(\rho, \phi, z)$ (see Fig. 1A). The index $n \in \{1,2\}$ is used to distinguish between the material that contains the dipole ($n=1$) and the material on the opposite side of the interface, containing no dipole ($n=2$). The distance of the dipole from the interface $d$ is defined to be positive when the dipole is located above and negative if it is located below the interface on the $z$-axis. $k_{SPP} = k_0\sqrt{(\epsilon_1\epsilon_2)/(\epsilon_1+\epsilon_2)}$ is the SPP propagation constant, where $\epsilon_1$ and $\epsilon_2$ are the frequency dependent, complex dielectric constants in the materials containing the dipole and not-containing the dipole respectively. The constants $k_{z,n}^2 = k_0^2 \epsilon_n - k_{SPP}^2$ for $n \in \{1,2\}$ determine the exponential decay of the SPP mode away from the interface, with the root chosen such that $\text{Im}\{k_{z,n}\} > 0$ and $n$ given according to the location of the dipole as before. $M$ is a coupling constant given by $M = -\frac{1}{2}\left(\frac{k_{z,1}k_{z,2}}{\sqrt{2\pi\epsilon_0}}\right)\left(\frac{k_{z,1}\epsilon_2 - k_{z,2}\epsilon_1}{\epsilon_1^2 - \epsilon_2^2}\right)\exp(-i\pi/4)$. The vector $\vec{v} = \hat{\rho} - \text{sgn}(z)(k_{SPP}/k_{z,n})\hat{z}$ describes the polarization of the SPP field. The angular dependence of the field is captured by the term in the second line of Eq. 2, where we recognize the angular dependence of the in- and out-of plane dipoles shown in Figs. 1B and C. The time-averaged SPP power radiated per unit angle in-plane $\partial P/\partial \phi$ is given by:[26]

$$\frac{\partial P}{\partial \phi} = 2\text{Re}\left\{\left(\frac{k_{SPP}}{\mu\omega}\right)\left[\left(\frac{k^*_{z,1}}{\text{Im}\{k_{z,1}\}k_{z,1}}\right)\left[1+\left(\frac{k_{SPP}}{k_{z,1}}\right)^2\right]^* + \left(\frac{k^*_{z,2}}{\text{Im}\{k_{z,2}\}k_{z,2}}\right)\left[1+\left(\frac{k_{SPP}}{k_{z,2}}\right)^2\right]^*\right]\right.$$
$$\left. \times \frac{\exp\left[-2\left(\text{Im}\{k_{SPP}\}\rho + \text{Im}\{k_{z,1}\}|d|\right)\right]}{|k_{SPP}|}|M|^2\,\Phi(\phi)\right\} \quad (0)$$

Where $\Phi(\phi)$ is the angular dependence to the SPP radiation pattern of a general dipole:

$$\Phi(\phi) = \left| p_x\cos\phi + p_y\sin\phi + p_z\left(\text{sgn}(d)\frac{k_{SPP}}{k_{z,1}}\right)\right|^2 \quad (0)$$

For general values of $\vec{p}$ and $k_{SPP}/k_{z,1}$, the function $\Phi(\phi) \neq \Phi(\phi+\pi)$, that is, the angular distribution of power radiated in form of SPPs is asymmetric. The above expression for $\Phi(\phi)$ can be rewritten into components that are symmetric and anti-symmetric upon inversion about the origin:

$$\Phi(\phi) = \overbrace{|p_x|^2|\cos\phi|^2 + |p_y|^2|\sin\phi|^2 + |p_z|^2|\Gamma|^2 + 2\text{Re}\{p_x p^*_y\}\cos\phi\sin\phi}^{\text{symmetric under }\phi\to\phi+\pi}$$
$$+ \underbrace{2\,\text{sgn}(d)\text{Re}\{(p^*_x\cos\phi + p^*_y\sin\phi)p_z\Gamma\}}_{\text{antisymmetric under }\phi\to\phi+\pi} \quad (5)$$

Where we have defined $\Gamma \equiv k_{SPP}/k_{z,1}$. Eq. (5) shows that $\Phi(\phi)$ always has a symmetric component, but will also have a non-vanishing anti-symmetric component if the dipole has components that oscillate both in- and out-of-plane and one or both of the following criteria are satisfied: (I) $\Gamma$ has a real component and/or (II) the in-plane components of the dipole oscillate with a non-vanishing phase relative to $p_z$, excluding a phase shift of exactly $\pi$.

Criterion (I) is fulfilled whenever losses (or gain) are present in the materials, which gives rise to imaginary components of the dielectric constants of the materials. If all dipole components oscillate in phase (or with a $\pi$-phase shift), the dipole oscillates along a linear axis that is tilted with respect to the interface. In this case, the extinction ratio (i.e. the ratio of the powers emitted towards the two opposing sides of the dipole) is maximal for a tilting angle of the dipole of $\alpha = \arctan|\Gamma|$ (cf. Eq. 1) and bounded by:

$$\max_{\phi \in [0, 2\pi]} \left\{ \frac{\Phi(\phi)}{\Phi(\phi + \pi)} \right\} = \frac{|\Gamma| + |\operatorname{Re}\{\Gamma\}|}{|\Gamma| - |\operatorname{Re}\{\Gamma\}|} \tag{6}$$

where maximal extinction (and emission) occurs in the direction $\phi = \beta$ if $\operatorname{Re}\{\Gamma\} > 0$ and $\phi = \beta + \pi$ if $\operatorname{Re}\{\Gamma\} < 0$. For most practical situations, the imaginary part of $\Gamma$ is about an order of magnitude larger than the real part and the asymmetry of the emission by a tilted linear dipole is hence small. In the perfectly lossless case, $\operatorname{Re}\{\Gamma\} = 0$ and there is no asymmetry.

Criterion (II) entails that the oscillation of the dipole corresponds to rotational motion along an elliptical or circular trajectory rather than an oscillation along a linear axis. In free space, the emission of such dipoles is elliptically or circularly polarized light along the axis of rotation. Fig 2A shows that the SPP radiation pattern of a rotating dipole in the vicinity of a real metal, i.e. the general case when both criteria (I) and (II) apply, can be quite complicated.

Allowing for rotating dipoles, the extinction can, in contrast to the case of the tilted linear dipole, be made perfect. This case is particularly interesting for the efficient steering of SPPs[14]. It is possible to engineer such a unidirectionally emitting dipole by adjusting the in- and out of plane components such that $\alpha = \arctan(|\Gamma|)$ and $\delta = \gamma = \arctan(\operatorname{Im}\{\Gamma\} / \operatorname{Re}\{\Gamma\})$, again using the notation defined in Eq. 1. $\Phi(\phi)$ then reduces to:

$$\Phi(\phi) = |p_0|^2 \cos^2\left[\arctan(|\Gamma|)\right] |\Gamma|^2 \left[\cos(\phi - \beta) + 1\right]^2 \tag{7}$$

In this case, the emission is maximal in the direction $\phi = \beta$ while it fully vanishes in the opposite direction $(\phi = \beta + \pi)$. Fig. 2B shows the SPP fields launched by such a dipole, emitting unidirectionally along the $x$-axis. In this ideal case, 92.4% of the power is emitted in the $x > 0$ half-space. The direction of unidirectional emission can be continuously tuned in-plane by changing $\beta$, or inverted by changing the phases $\delta \to \delta + \pi, \gamma \to \gamma + \pi$. In Fig. 3 we show radiation patterns of a dipole with $\alpha = \arctan(|\Gamma|)$ for various phase shifts $\delta$, $\gamma$ and angles $\beta$, showing the tunable directional launching of SPPs. Following the completion of this work,[27] weakly unidirectional SPP excitation due to this effect was experimentally realized by mimicking a rotating dipole chain with a slit in a metal surface that was illuminated with circularly polarized light at grazing incidence. [24]

In conclusion, we have shown that the SPP radiation patterns of a point-dipole near a metal surface are generally asymmetric and exhibit a variety specific to their mode of dipole oscillation. A careful measurement of the SPP radiation pattern, for example using leakage radiation microscopy (LRM)[1,3,24,25], can be used to make strong inferences regarding the nature of the emitter, such as the transition dipole moment of a fluorophore or the induced dipole moment in a subwavelength scatterer[1,18-20]. This may be particularly relevant for surface plasmon-coupled emission (SPCE)-type experiments in biochemical sensing, where the directionality of the out-of-plane leakage of SPPs already plays a central role, but where the in-plane angular dependence has not yet been investigated[2,3]. If, on the other hand, the properties of the emitter are under the control of the experimenter, the directionality of SPP emission can be engineered. Examples of such emitters are quantum wells with spin-polarized electrons, which give rise to circularly polarized emission across the interband heavy-hole transition[28] and aligned luminescent materials[29,30]. Engineered scatterers, such as optical antennas[31], also offer control over the induced dipole moment that eventually couples to SPP modes[24] Certain single photon sources, such as NV-centers in diamond, have well-understood polarized luminescence properties[32] that could in principle result in asymmetric emission of SPPs. The selective excitation of Zeeman split transitions in atoms near a metal surface may also constitute an avenue to create circularly polarized nearfield sources. Another approach to engineering SPP radiation patterns may consist in engineering the properties of the plasmonic mode (hence the factor $\Gamma$) by changing the environment or frequency of the emitter. The effect of the change in material properties on the SPP radiation pattern may then be used for novel sensors or beam steering applications. We note that asymmetry can arise also from higher order corrections to the dipole approximation, such as magnetic dipole and electric quadrupole transitions, though those emission channels are typically much weaker and beyond the scope of this paper.

**Acknowledgements:** This research was supported by the Air Force Office of Scientific Research under grant FA9550-12-1-0289. The authors express thanks to S. J. Byrnes, H. Galinski (Harvard SEAS) and A. Belyanin (Texas A&M Univ.) for helpful discussions.

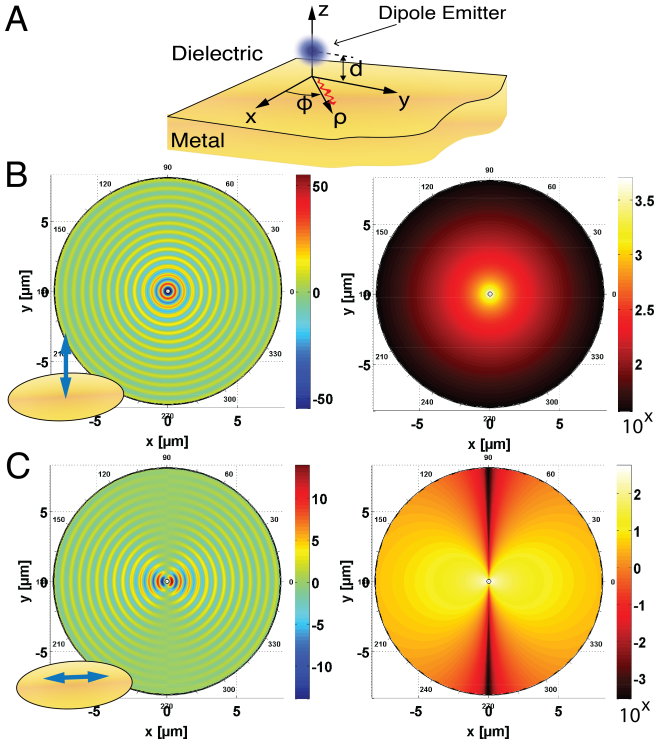

**Figure 1:** (Color online) **(A)** An SPP-emitting dipole is situated on the z-axis at a distance d from a metal/dielectric interface in the xy-plane. **(B)** Left: The $z$-component of the instantaneous SPP electric field in $[V/m]$ generated by a dipole in air that oscillates out-of-plane with respect to a gold surface ($\epsilon_1 = 1; \epsilon_2 = -9.9 + 1.0i$) at a frequency corresponding to a vacuum wavelength of 652nm. The dipole is situated a distance $d = 50$nm above the interface and has strength of $p_0 = 1$D ($1\text{Debye} = 3.3356 \times 10^{-30}$Cm). The field is evaluated on the air-side ($n = 1$), at a distance of $z = 5$nm above the interface. Right: Plot of the time-averaged electric field intensity of the SPP field at $z = 5$nm above the interface ($\left\langle \left|\vec{E}\right|^2 \right\rangle_t$) in $[V^2/m^2]$. The color scale is in powers of 10 to enhance visibility. **(C)** Left: The z-component of the instantaneous SPP electric field at $z = 5$nm created by a dipole as in (A), oscillating in-plane with the metal surface in $[V/m]$. Right: plot of the time-averaged electric field intensity at $z = 5$nm for the parallel dipole in $[V^2/m^2]$. The plots take near-field effects into account, which add small perturbations in the immediate vicinity of the source.[26]

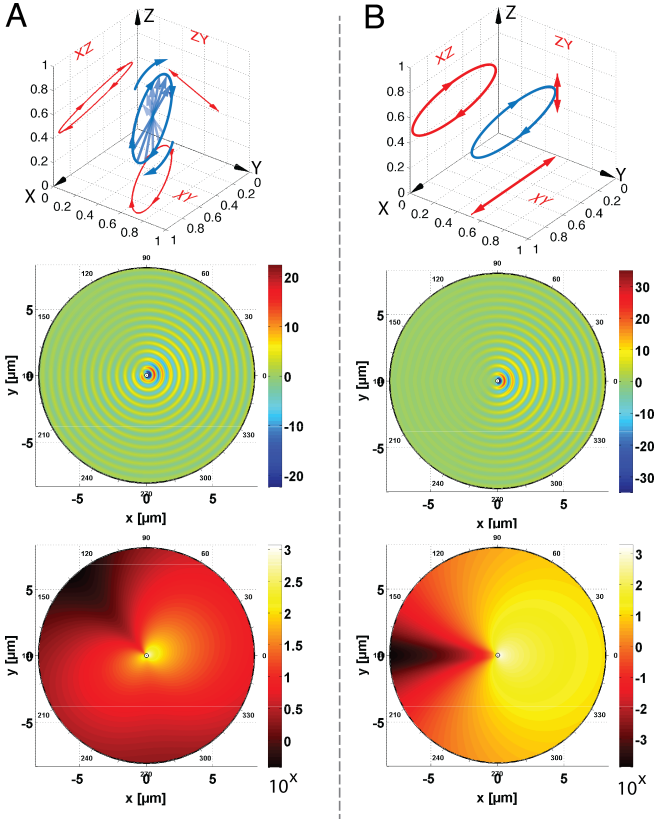

**Figure 2:** (Color online) **(A)** Top: the time-evolution of a dipole (blue arrows) with $\alpha = 0.44\pi, \beta = -0.15\pi, \delta = -0.83\pi, \gamma = 0$ in arbitrary units. The dipole rotates, tracingout an ellipsis (blue) in 3D space. The projections of the trajectory of the dipole on the $xz$, $zy$ and $xy$ plane are also shown (red). Center: The z-component of the surface plasmon polariton (SPP) electric field launched by the dipole at an air/gold interface, evaluated at $z = 5$nm above the gold interface and in $[V/m]$. The dipole emits at a frequency corresponding to a vacuum wavelength of 652 nm. Bottom: plot of the time averaged electric field intensity at $z = 5$nm. In contrast to the fields plotted in Figs. 1B and C, the radiation field is clearly asymmetric about the origin. **(B)** Top: the time-evolution of a dipole (blue) of strength $p_0 = 1$ that rotates out of plane at the same frequency with parameters chosen as $\alpha = \arctan(|\Gamma|) \approx 0.402\pi$, $\beta = 0$, $\delta = \gamma = \arctan(\mathrm{Im}\{\Gamma\}/\mathrm{Re}\{\Gamma\}) \approx -0.517\pi$. This choice ensures that the dipole launches SPPs unidirectionally in the $+x$ direction when placed in air above the metal surface. Center: The $z$-component of the SPP electric field (in $V/m$) generated by the unidirectionally emitting dipole. Bottom: Plot of the electric field intensity at $z = 5$nm in $[V^2/m^2]$. The plots take the nearfield of the dipole into account,[26] hence the extinction can be seen to be imperfect directly to the left of the dipole. The values used to specify geometry, material properties and dipole strength in (A) and (B) are the same as used in Figs. 1B and C

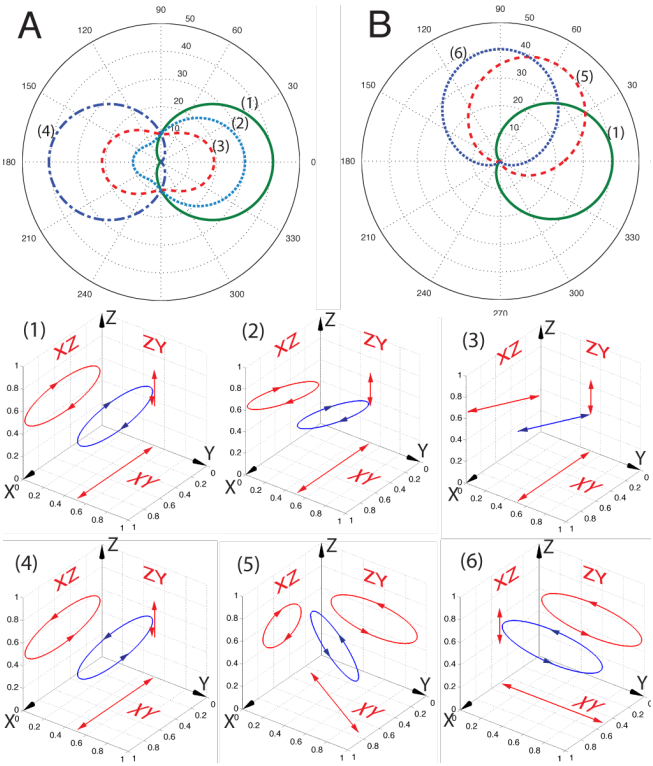

**Figure 3:** (Color online) **(A)** The SPP radiation pattern of a dipole in air above a gold surface with $\alpha = \alpha_{uni} = \arctan(|\Gamma|) \approx 0.402\pi$ and $\beta = 0$ for various values of the phase $\delta$, showing continuous tuning of the emission from right-unidirectionality to left-unidirectionality via intermediate extinction ratios (in arbitrary units). The phases are (1) $\delta = \delta_{uni} = \arctan(\operatorname{Im}\{\Gamma\}/\operatorname{Re}\{\Gamma\}) \approx -0.517\pi$ (right unidirectional), (2) $\delta = \delta_{uni} + 0.33\pi$, (3) $\delta = 0$ and (4) $\delta = \delta_{uni} + \pi$ (left-unidirectional). While out-of-plane rotating dipoles lead to a highly asymmetric emission of SPPs, a tilted linear dipole will also create an asymmetric pattern, as shown in curve (3) due to optical losses in the gold. The strength of the asymmetry increases with the losses in the material. **(B)** The SPP radiation patterns of a unidirectionally emitting dipole ($\alpha = \alpha_{uni}, \delta = \gamma = \delta_{uni}$) for several values of $\beta$ (in arbitrary units), showing how the direction of unidirectional emission can be moved in plane by rotating the axis of rotation of the dipole about the $z$-axis. The values of $\beta$ are: (1) $\beta = 0$, (2) $\beta = \pi/3$ and (6) $\beta = \pi/2$. The dielectric constants, dipole strength and frequency are the same as in Fig. 1 and 2. The motion of the dipoles is plotted in the center and bottom parts of the figure in diagrams labeled from (1) to (6). The motion of the dipole is shown in blue, with the projections of the trajectory on the $xz$, $zy$ and $xy$ plane shown in red (arbitrary units).